# Shot-Noise limited Time-encoded (TICO) Raman spectroscopy

**Sebastian Karpf[1,4], Matthias Eibl[1,2], Wolfgang Wieser[3], Thomas Klein[3], Robert Huber[1,2,*]**

[1]*Lehrstuhl für BioMolekulare Optik, Fakultät für Physik, Ludwig-Maximilians-Universität München Oettingenstr. 67, 80538 Munich, Germany*
[2]*Institut für Biomedizinische Optik, Universität zu Lübeck, Peter-Monnik-Weg 4, 23562 Lübeck, Germany*
[3]*Optores GmbH, Gollierstr. 70, 80339 Munich, Germany*
[4]*Department of Electrical Engineering, University of California Los Angeles, Los Angeles, California, United States of America*
[*]*robert.huber@bmo.uni-luebeck.de*

## Abstract:

Raman scattering, an inelastic scattering mechanism, provides information about molecular excitation energies and can be used to identify chemical compounds. Albeit being a powerful analysis tool, especially for label-free biomedical imaging with molecular contrast, it suffers from inherently low signal levels. This practical limitation can be overcome by non-linear enhancement techniques like stimulated Raman scattering (SRS). In SRS, an additional light source stimulates the Raman scattering process. This can lead to orders of magnitude increase in signal levels and hence faster acquisition in biomedical imaging. However, achieving a broad spectral coverage in SRS is technically challenging and the signal is no longer background-free, as either stimulated Raman gain (SRG) or loss (SRL) is measured, turning a sensitivity limit into a dynamic range limit. Thus, the signal has to be isolated from the laser background light, requiring elaborate methods for minimizing detection noise. Here we analyze the detection sensitivity of a shot-noise limited broadband stimulated time-encoded Raman (TICO-Raman) system in detail. In time-encoded Raman, a wavelength-swept Fourier Domain Mode Locked (FDML) laser covers a broad range of Raman transition energies while allowing a dual-balanced detection for lowering the detection noise to the fundamental shot-noise limit.

## Introduction:

Optical spectroscopy can yield a wealth of information for biomedical imaging and diagnostics. In fluorescent imaging, for example, a set of diverse fluorescent dyes can help to visualize different functional sites and dynamics at the sub-cellular level. The high optical resolution and the accessibility of the information play a crucial role in the study of biological research and diagnostics. Raman spectroscopy promises to provide this diverse information by inelastic light scattering, without the need of staining the sample. The big advantage of this label-free operation is that it is non-invasive and does not risk altering the biological function of the sample. However, for in vivo application of Raman spectroscopy high speed and specificity is needed, so that prominent candidates employ non-linear Raman techniques for signal level enhancements [1-16].

We previously reported on a new technique for non-linear, stimulated Raman scattering (SRS), called time-encoded Raman technique (TICO-Raman) [17]. In TICO-Raman, the SRS effect is encoded and detected in time-domain rather than in frequency-domain. Latest developments in telecommunication-driven, high speed analog-to-digital converter (ADC) technology promise faster acquisition speeds in Raman spectroscopy [15]. Broadband spectra are recorded by encoding the Raman spectral coverage in time by a wavelength-swept Fourier Domain Mode Locked (FDML) laser[18]. An unambiguous time-to-Raman energy mapping is achieved where the Raman signal height is measured as intensity change of the FDML laser. A unique set of measures were introduced that result in a maximally reduced noise contribution to the SRS spectra, where the ultimate shot-noise limit was reached. In this paper we show the signal extraction process in detail and visualize the two-stage, analogue and digital balanced detection mechanism for shot-noise limited performance.

## Materials and Methods:

The system comprises two home-built lasers, an FDML Raman probe laser and a Master-Oscillator Fiber Power Amplifier (MOFPA) laser as programmable Raman pump laser (Figure 1). The home-built MOFPA laser incorporates a fast electro-optical modulator (EOM) which allows for active modulation of the pulses. The system employs a pulse pattern that probes all possible Raman transitions via the TICO-Raman technique [17]. The detection comprises a balanced photodetector and a fast analog-to-digital converter card (ADC). The pump pulse length is digitally set to match the analog bandwidth of the detection system. As the SRS effect is an instantaneous effect, the stimulated Raman gain (SRG) signal follows the time characteristic of the Raman pump pulse. By matching the pump pulse length to the detection bandwidth the SRG effect height is fully and optimally digitized.

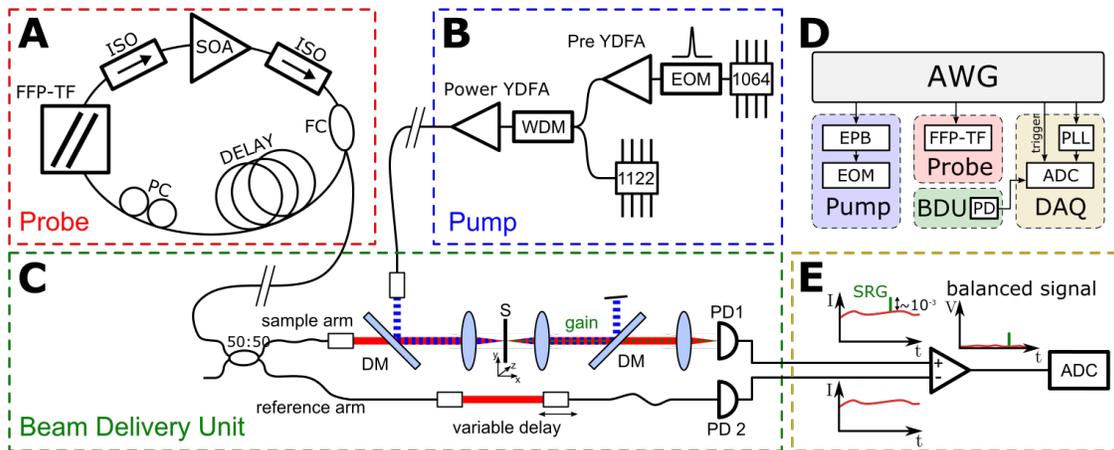

Figure 1 The schematic of the shot-noise limited TICO-Raman spectroscopy system. (A) The FDML probe laser comprises a fiber Fabry-Pérot tunable filter (FFP-TF), optical isolators (ISO) for uni-directional lasing, a semiconductor optical amplifier (SOA), a fiber-coupler (FC) output, a resonant time delay, and a polarization controller (PC) paddle. (B) The pump laser is a home-built Master Oscillator Fiber Power Amplifier (MOFPA) with additional Raman-shifter. It consists of a narrowline laser diode at 1064nm, followed by a fast electro-optical amplitude modulator (EOM). This EOM is used to digitally modulate the TICO pulse pattern and to adjust the pulse length to the detection bandwidth. The pulses are amplified by ytterbium-doped fiber amplifiers (YDFA). A wavelength-division multiplexer (WDM) can be used to couple in additional 1122nm seed light for Raman-shifting inside the double-clad YDFA power amplifier. (C) The pump and probe lasers are collimated and combined by dichroic mirrors (DM) before being focused onto the Raman active sample (S). After recollimation, the pump light is blocked by additional dichroic filters and detected on a photodiode (PD1). A second photodiode (PD2) is used to detect reference light for differential, balanced detection. The differential transimpedance amplifier (E) subtracts FDML offset light and common mode-noise, so only intensity changes induced in the Raman-active sample are digitized by the data acquisition (DAQ). (D) The whole system is synchronized by an arbitrary waveform generator (AWG).

The process from SRG signal detection to Raman spectra generation is demonstrated using SRG data of neat benzene (Figure 2). The SRG signal is recorded as power change of the probe laser power. We employ a balanced detector with differential amplifier to remove the DC offset and only detect relative power changes. This technique has two advantages: a) laser noise can be reduced by the common mode rejection of the photoreceiver and b) the range of the ADC can be utilized optimally to digitize the power change and no bits are lost to digitizing the DC offset. The first, analog balancing step consists of the differential amplification of the FDML probe light (rainbow color, left). The probe laser is split in equal parts, one with pump laser interaction and the other without, thus serving as reference light. This analog balancing step additionally leads to common mode noise rejection. For highest possible common mode noise rejection ratio, it is crucial to adjust the two arm lengths to match as closely as possible in order to match the phase of any common electronic noise. Considering the bandwidth of our detection of 400MHz, the fastest electronic signal may have a period of 2.5ns. This corresponds to 2Pi phase. For a differential balancing suppression of a factor of 1000, the phase difference of the two arms should be less than one thousandth of Pi/2 (linear approximation), i.e. less than 625fs. In free-space this corresponds to a distance travelled by light of 3E8*625E-15≈190μm. We employed a micrometer screw in the reference arm to adjust for this length. Furthermore, the power was adjusted to carefully match the two arm lengths' powers.

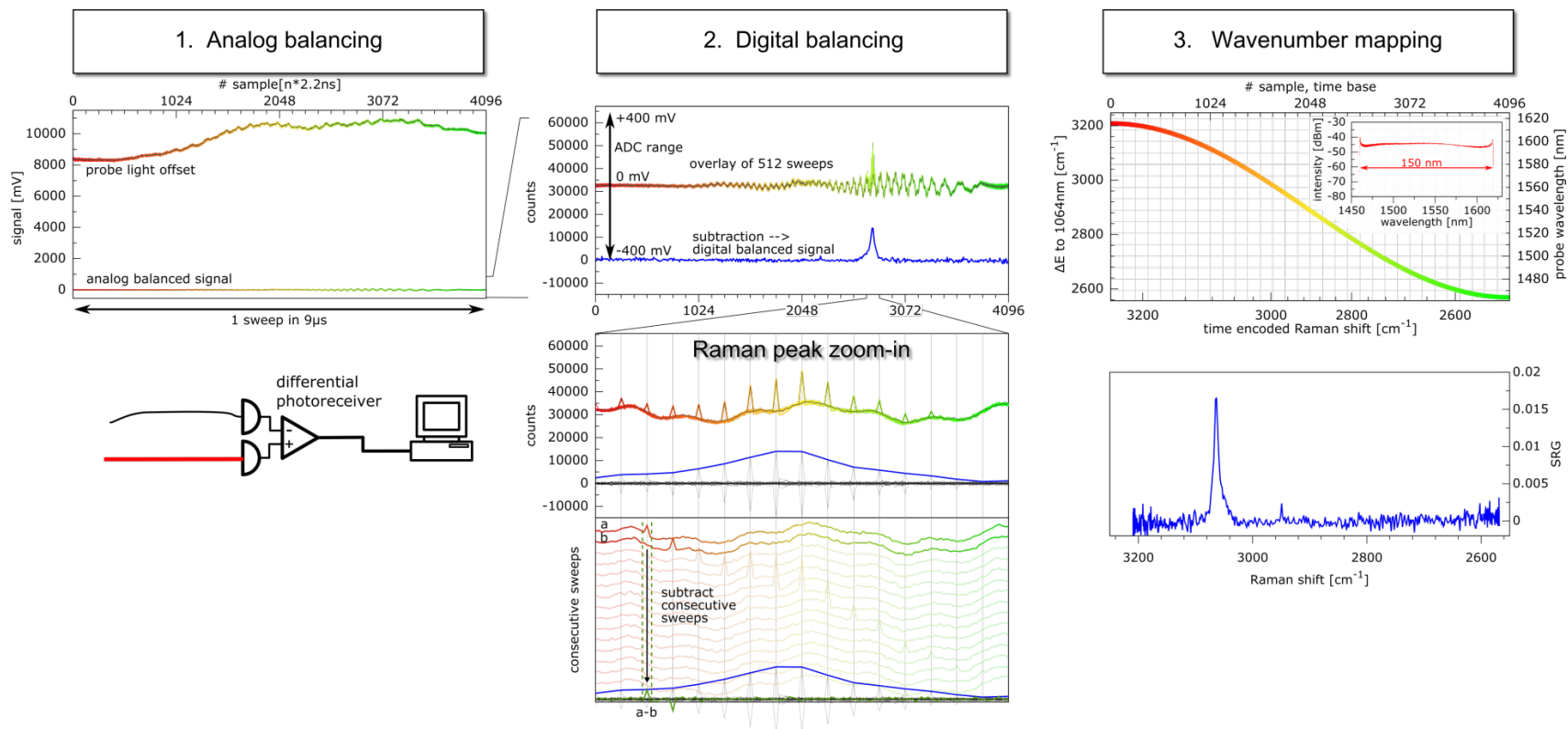

Figure 2 Raman spectra extraction by two-stage balanced detection, leading to an SRG spectrum of benzene. Left: The first analog balancing step employs a balanced, differentially amplified photoreceiver to remove the probe light offset. Only power changes in the sample arm remain (cf. Figure 1). Middle: The analog-to-digital converter card (ADC) is used to synchronously sample the FDML probe laser sweeps. The 800mV range is sampled at 14-bit depth resolution. The lower part shows a zoom-in on a part where SRG signals occur. The vertical lines represent the positions of the pump pulses following the TICO-Raman excitation technique [17]. The SRG signals are clearly visible and fully recorded as ensured by the synchronized detection (see text). The second, digital balancing step subtracts consecutive sweeps to further lower the noise level to the shot-noise limit. Only pump induced intensity changes remain. This step makes use of TICO-Raman pattern, where pulse positions increase from sweep to sweep. Thus, through subtraction any artifacts due to chromatic imperfections are removed. Right: The last step assigns wavenumber values to the Raman spectrum. First the spectrum of the FDML laser (inset) is recorded on an optical spectrum analyzer (OSA). Then, the wavelength-to-time mapping of the FDML laser is calculated and from that the energy difference to the pump laser is determined. The result is accurate Raman shift energies for each spectral point. Shown in blue is the final, shot-noise limited TICO-Raman spectrum of benzene, acquired in 9ms.

A zoom into the subtracted, amplified signal is presented in the second column of Figure 2. Shown are 512 consecutive sweeps of the FDML laser centered at 0mV. They only contain intensity changes occurring in the sample arm. First, this shows the very good sweep-to-sweep correlation of the FDML laser. Even though a residual modulation is visible, which stems from chromatic imperfections between the sample arm and the reference arm, it is clear that this spurious signal is present on all 512 sweeps and can thus be subtracted digitally. Hence we call this second step digital balancing. First, the 800mV range of the ADC (Alazartech ATS9360) samples the balanced intensity changes of the FDML probe laser at 12-bit depth resolution. Now, consecutive sweeps are subtracted (lower graph, middle column) to remove any remaining intensity differences between sample and reference arm. The SRG signals are not subtracted, as the pump pulse position increases stepwise for consecutive sweeps in the TICO-Raman technique (Figure 2, lower middle column). The effect of this digital balancing step is that only pump intensity changes remain. The result is shown in blue and already represents the SRG Raman spectrum. This step ultimately brings the noise level down to the fundamental shot-noise limit, which lies at around $3.6 \times 10^{-4}$ of the probe laser power of 2mW. To assess the overall advantage of the dual balancing method we will compare the dynamic range increase. First, the DC-offset of the FDML probe laser light lies around 10V (cf. Figure 2). Calculating the effective number of bits (ENOB) for the 12-bit ADC using ENOB=(SNR-1.76dB)/6.02 and 57dB SNR (manufacturer datasheet) gives ~9-bits ENOB resolution for the used Alazartech ATS9360. Thus, the full 800mV scale is divided into 512 units. Thus, with DC-offset a measurement sensitivity at the shot-noise level of $10^{-4}$ is not possible. Consequently, our first analogue balancing step reduces the required voltage range from 10V to 800 mV, i.e. a factor of 12.5. Hence, the effective voltage scale division after this step is 512*12.5=6400, which increases the dynamic range such that digitization can distinguish relative changes of $1/6400=1.56 \times 10^{-4}$, i.e. over a factor of 2 smaller than the relative shot-noise. In the second balancing step, following the digitization step, the aim is to reduce the contribution of any digitized noise or residual modulation. By subtracting consecutive sweeps the digital balancing steps acts as a filter which increases the dynamic range. Before this digital balancing step, the residual modulation height is about 1/7-th of the 800mV range, i.e. ~120mV. After digital balancing, all of the additional noise and modulation contributions are cancelled out and the spectrum is recorded at its shot noise-limit at $3.6 \times 10^{-4}$ relative signal height. In the last step of the TICO-Raman technique, Raman transition energies are assigned to the recorded SRG values. This is shown in the right column of Figure 2. The well-defined wavelength-to-time characteristic of the FDML allows mapping energy differences between pump laser and FDML sweep by recording a spectrum of the FDML (see also supplementary information of [17]). In combination, these steps enable the recording of Raman spectra of SRG values with accurate Raman transition energy values.

Precise timing and synchronization is crucial for the TICO-Raman technique. The whole system is driven by an arbitrary waveform generator (AWG), which electronically drives the Fabry-Pérot-Filter in the FDML laser. A second channel of the AWG is used to generate the TICO-Raman pump pulse pattern [17]. A third channel is used to trigger the acquisition and to provide a synchronized sample clock to the ADC card by means of a phase-locked loop (PLL). This electronic synchronization of the whole system requires an initial setting of the phases according to the relative time delays. In order to correctly time the start of the FDML probe laser sweep, an objective glass slide is inserted in the free-space beam. This ~1mm thick glass creates interference fringes on the FDML light (see Figure 3). These interference fringes can be used to adjust the phase of the FDML sweep to the start of the TICO Raman pattern. When the probe and pump laser are synchronized, the glass plate is removed. Now, the detection is synchronized. This step requires more accurate fine-tuning as the 1.8ns pump pulse length needs to be perfectly timed with respect to the ADC sample clock. To this end, the pulses are accurately timed by employing a home-built electronic pulse board (EPB) by adjusting a delay with respect to the trigger. The delay is set digitally (via USB) from 0ns to 10ns with 0.01ns accuracy. Then, the pump wavelength filters are removed, the FDML laser blocked, and the pump pulses are recorded on the ADC. Low power pump pulses without amplifiers are employed, so a saturation of the photodetector is avoided. The digitized pulse height is set maximal by tuning the time delay on the EPB. This ensures full effect height recordings of the instantaneous SRG effect (following the pump pulse timing). Once the delay on the EPB is set, the pump blocking filters are put back and the FDML laser is unblocked. Finally, a DC-coupled version of the FDML probe sweep is recorded, such that the SRS values can be transformed to SRG values as intensity changes of the probe light power. Therefore, a 99/1 fiber coupler after the FDML laser output (before the beam delivery unit) provides a 1% output, which is plugged into a DC-coupled photodiode of known transimpedance gain. This curve is recorded on the second channel of the ADC and adjusted by the power and transimpedance difference of the sample path (cf. Figure 2, left).

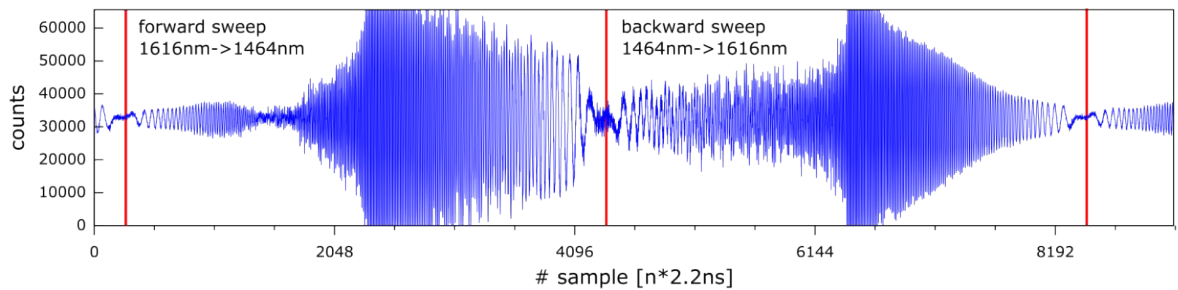

Figure 3 Strategy for time synchronization of the FDML sweep to the TICO-Raman technique. Interference fringes from the FDML probe light are generated by inserting an objective glass slide (~1mm thick) into the beam. These fringes are used to determine the phase of the FDML sweep. This phase can then be adjusted on the arbitrary waveform generator (AWG), such that the first pulse (vertical line) of the TICO-Raman pattern [17] coincides with the lowest wavelength of the sweep. This is crucial for a correct time-to-wavenumber mapping in the TICO-Raman technique.

## Results & Discussion:

To assess possible noise limitations it is important to test experimentally how the noise level decreases upon averaging and how close it lies to the ideal square-root behavior. To test this, we acquired a series of spectra with different averaging values. Figure 4 shows the TICO-Raman spectrum of benzene, from single acquisition up to 10.000-times averaging. The recording time for a single spectrum was 9ms. Upon averaging, two smaller Raman transitions around 3180$cm^{-1}$ become clearly visible. For quantitative analysis of the improvement, we analyzed the signal-to-noise ratio (SNR) by considering the SRG signal height of 1.69 x $10^{-3}$ for the 3063$cm^{-1}$ peak of benzene and dividing by the standard deviation in regions where no Raman transitions occur. We found, that the SNR increases from 34 to 508 when considering the spectral region from 2800$cm^{-1}$ to 2900$cm^{-1}$ (region I, 52 samples). However, when comparing to region II (30 samples) the SNR increases up to 1594. This 3-fold difference can be explained by an underlying baseline in the spectrum, which supposedly stems from cross-phase modulation (XPM). The effect shows a chromatic dependency, increasing with decreasing wavenumber values. This linearly increasing baseline adds to the standard deviation in region I. However, in region II the FDML sweep is close to its turning point, where the sweep direction changes. Therefore, in region II almost no change in wavelength occurs. Hence, in region II the effect is nearly constant and has no effect on the standard deviation, so the noise decreases close to the expected value from the ideal square-root behavior (Figure 4, right).

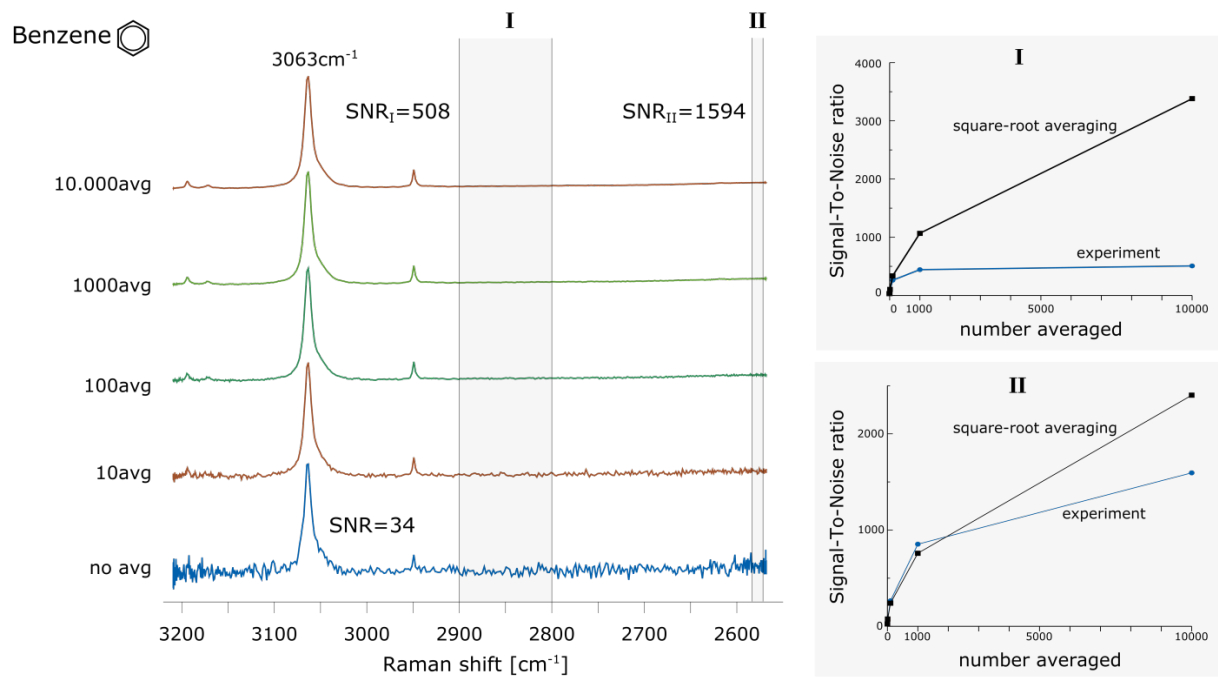

Figure 4 The result of averaging on the TICO-Raman spectrum of benzene. The lower, blue spectrum is the non-averaged spectrum. The signal (3060cm$^{-1}$ peak) is 34-times higher than the standard deviation of the areas in I or II. Averaging leads to much increased Raman spectra and new spectral features become visible. The different spectra were vertically offset for clarity. The increase in signal-to-noise ratio (SNR) over the amount of averaging is plotted on the right, where two different regions of the spectrum were used to calculate the noise. Region I shows a higher standard deviation due to the contribution of an additional effect – we assume cross-phase modulation (XPM, see text). In region II, this effect is constant and does not enter the standard deviation, leading to an almost optimal square-root increase in SNR.

The effectiveness of the signal detection is best shown by considering the signal to noise ratio (SNR). The SNR depends on both the noise level, which is shot-noise limited here as explained, but also on the signal level. In order to fully record the intensity change, it is crucial to fully resolve the time response of the probe laser. We used pump pulses which directly correspond to the analog bandwidth of the detection system, such that the whole signal due to SRG interaction is digitized. This direct detection makes the recovery of absolute transmission changes straight forward and it is possible since we employ nanosecond lasers. The fact that the light is delivered through a single mode fiber with a defined beam profile additionally helps any quantitative measurements of Raman cross-sections.

Our SRG signal levels are in very good agreement with values previously reported in literature [19]. Owyoung et al. report a value of $1.75 \times 10^{-5}$ for the intense Raman transition of benzene at 992cm$^{-1}$ using 50mW of pump power. We can compare this value to our recorded SRG intensity of the 3060cm$^{-1}$ line of benzene (Figure 2). Taking into consideration the quadratic dependency of the pump and probe wavelengths (factor 4.5), the employed peak pump power of 1.6kW and a factor of 6.9 for the lower spectral cross-section of the 3060cm$^{-1}$ line compared to the 992cm$^{-1}$ line, one calculates an expected SRG signal of 1.81 x$10^{-2}$. Our measured value of 1.69 x $10^{-2}$ agrees very well and shows the advantage of using the nanosecond approach, especially if quantitative measurements are desired. One interesting new approach is the application of ultra-short pulses in combination with Photonic Time-Stretch, effectively enhancing the detection bandwidth and sampling in time-domain [15].

We also found imperfections with the lenses when using wavelengths in the extended Near-Infrared (exNIR). The current study employed standard fused glass lenses (cf. [17]). Chromatic aberrations currently lead to imperfections of the recorded SRG values over the whole spectral range. If quantitative measurements are to be performed it is important that better engineered exNIR objectives are employed ensuring same spot sizes for pump and probe lasers along the broad spectral coverage.

## Conclusion:

We presented strategies for a shot-noise limited system for broadband SRS spectroscopy by employing a wavelength swept FDML laser and a time-domain based acquisition. Measures were described in detail how to effectively reduce the detection noise down to the fundamental shot-noise limit. These include one analogue and one digital balancing step. The nanosecond pulses of our pump laser match the detection bandwidth, allowing a full recovery of the SRG signal height. This capability can lead to precise quantitative measurements while effect heights are non-linearly enhanced due to SRS. The laser output from a single mode fiber with a well defined mode field diameter further aids quantitative analysis. Future investigations will focus on improving the sensitivity of the system through enhancing the SNR. To this end, the relative shot-noise will be lowered by employing higher, pulsed probe laser powers. Another promising approach is increasing the detection bandwidth, thereby gaining linearly in effect height while the shot-noise contribution only scales with the square-root. Overall, this powerful new tool for SRS spectroscopy can propel a biological and medical application in label-free imaging and diagnostics.

## Acknowledgements:

We thank A. Vogel from the University of Lübeck. We further acknowledge funding from the European Union project ENCOMOLE-2i (Horizon 2020, ERC CoG no. 646669) and German Science Foundation (DFG projects HU1006/6).

## Competing Financial Interests

The Ludwig-Maximilians-University has filed patent applications based on this work. W.W., T.K. and R.H. have financial interests in OptoRes GmbH commercializing FDML lasers.